# Climatic and Biogeochemical Effects of a Galactic Gamma-Ray Burst

Adrian L. Melott[1], Brian C. Thomas[1], Daniel P. Hogan[1], Larissa M. Ejzak[1], and Charles H. Jackman[2]

[1] Department of Physics and Astronomy, University of Kansas, Lawrence, Kansas, USA
[2] Laboratory for Atmospheres, NASA Goddard Space Flight Center, Greenbelt, Maryland, USA

## ABSTRACT

It is likely that one or more gamma-ray bursts within our galaxy have strongly irradiated the Earth in the last Gy. This produces significant atmospheric ionization and dissociation, resulting in ozone depletion and DNA-damaging ultraviolet solar flux reaching the surface for up to a decade. Here we show the first detailed computation of two other significant effects. Visible opacity of $NO_2$ is sufficient to reduce solar energy at the surface up to a few percent, with the greatest effect at the poles, which may be sufficient to initiate glaciation. Rainout of dilute nitric acid is could have been important for a burst nearer than our conservative "nearest burst". These results support the hypothesis that the late characteristics of the late Ordovician mass extinction are consistent with GRB initiation.

# 1. Terrestrial implications of gamma-ray bursts in our galaxy

Gamma-ray bursts (GRBs) are now determined to be at cosmological distances and by implication have enormous luminosity in gamma and X-ray photons [*Mészáros* 2001], and possibly cosmic rays as well [*Waxman* 2004; *Dermer and Holmes* 2005]. It has been recognized that they may constitute a significant threat to the Earth and any other life-bearing planets [e.g. *Thorsett* 1995; *Scalo and Wheeler* 2002]. Based on the latest rate information, it has been estimated [*Melott et al.* 2004; *Dermer and Holmes* 2005] that the Earth has been irradiated from a distance of order 2 kpc within our galaxy within the last Gy, approximately the time the Earth has possessed a significant ozone shield.

For some time, it has been known that high energy radiation may, through dissociation of $N_2$, create a variety of "odd nitrogen" compounds which lead to ozone depletion, making the atmosphere more transparent to solar UVB (290-315 nm) radiation. UVB is strongly absorbed by the DNA molecule and hazardous to life [e.g. *Cockell* 1999]. We have recently performed the first detailed simulation of the effect of a Milky Way galaxy GRB on terrestrial atmospheric chemistry [*Thomas et al.* 2005a,b]. While there is considerable variability of effect due to latitude and time of year, most of the probable parameter space of proximate bursts is characterized by significant ozone depletion lasting for years, at levels consistent with severe ecological damage. *Melott et al.* [2004] have argued that this may have been a cause of one or more of the terrestrial mass extinctions, and noted that the late Ordovician mass extinction 443 Mya was worse for surface or shallow-water organisms, which they characterize as consistent with UVB damage.

*Reid et al.* [1978] noted two other potentially important effects, which have been acknowledged [*Thorsett* 1995] but not yet treated quantitatively in subsequent discussions of GRB atmospheric ionization effects. $NO_2$, one of the primary compounds formed, absorbs strongly in the visible, giving it a brown cast. Such absorption could lower global temperatures. Also, rainout of dilute nitric acid ($HNO_3$) is one of the principal mechanisms of removal for the so-called "odd nitrogen" or "$NO_y$" compounds formed. This can potentially contribute large amounts of biologically active nitrogen to the biosphere. The results are unpredictable but may be major, since biota are typically nitrate-starved and highly responsive to supplementation [*Schlesinger* 1997].

A glaciation followed by an accelerated invasion of the land characterize the late Ordovician mass extinction, which has otherwise been suggested as consistent with GRB initiation [*Melott et al.* 2004]. In this *Letter* we explore the implications of our atmospheric computation for the terrestrial energy balance and nitrate deposition to further assess their possible connection to such events.

## 2. The computational model, its input parameters and resulting $O_3$ depletion

We have used the Goddard Space Flight Center 2D atmospheric model as extensively described elsewhere [*Douglass* et al. 1989; *Jackman* et al. 1990; *Considine* et al. 1994]. The two dimensions are latitude (18 bands) and altitude (58 bands). A lookup table is used for computation of photolysis [*Jackman et al.* 1996], and anthropogenic compounds are not included for our present purpose [see also *Thomas et al.* 2005b]. Winds and small scale mixing are included as described in *Fleming et al*. [1999]. We initially input the GRB radiation with a 10s step function burst divided into 1s timesteps, perturbing an atmosphere which resulted from a long-term equilibrium model. The results of this perturbed atmosphere were then re-introduced into the long-term model, run with a 1 day timestep for 20 years. The GRB spectrum was modeled as a typical event after *Band et al*. [2003], peaking at 187.5 keV. The total fluence was scaled to 100 kJ/m$^2$, corresponding to a conservative choice (2 kpc) of the nearest likely long-burst GRB beamed at the Earth in the last Gy [*Melott et al*. 2004; *Dermer and Holmes* 2005]. If sufficiently high-energy cosmic rays accompany GRBs [*Dermer and Holmes* 2005], a substantial additional effect (not included) from more penetrating muon secondaries is likely. Although the range of uncertainty in the energetics and distance of possible GRBs includes somewhat greater total fluence than 100 kJ/m$^2$, we have not modeled such events, as the large energy input exceeds the range of validity of our computational model.

Our model is empirical, based on the current Earth, with its continent and wind configuration. This introduces uncertainty, including a north-south asymmetry, because the northern and southern hemispheres of the current Earth are not mirror images. Therefore the model is not fully general, due to continental drift. We have tested the effect of this by doing a large ensemble of bursts at 4 seasons and 5 latitudes (+90, +45, 0, -45, and -90); full results are presented in *Thomas et al.* [2005b]. Our tests indicate that uncertainty in most results due to location and timing of unknown past GRB events is much greater than uncertainty due to continental drift. However, as discussed below, our rainout results are less generic

The purpose of this letter is to characterize the cooling effect and nitrate deposition resulting from a GRB that may reasonably be expected to have interacted with the Earth in the last Gy. Before about 0.4 Gya there was little presence of life on land [*Wellman et al.* 2003]. Among marine organisms, only surface floating or extremely shallow water dwellers could be directly affected by scattered GRB photons, since primarily UVB would reach the ground from the burst [*Smith et al.* 2004] and water absorbs UVB.

A large burst of ionizing radiation ionizes and dissociates molecules, leading to formation of the family of "odd nitrogen" compounds, most importantly NO and NO$_2$. These together catalyze the destruction of O$_3$. Despite the brevity of the initial burst, total O$_3$ depletion can easily reach global mean levels of 50%, tripling the UVB load. Even small increases in UVB, normally shielded by $O_3$, are known to be hazardous to phytoplankton and many other organisms [*Häder et al* 2003]. Bursts at very high latitudes may have the $O_3$ depletion (and other effects) largely confined to one hemisphere. When this is convolved with solar irradiance levels, the DNA damage is typically more severe at equatorial and/or mid-latitudes [*Thomas et al.* 2005b]. As UVB is absorbed by a few meters or 10s of meters of water, depending on turbidity and dissolved organics, organisms living near the surface, in shallow water, or with planktonic laval forms should be very hard-hit. The consistency of this with the Late Ordovician extinction pattern is beyond the scope of this Letter, but is discussed with references in *Melott et al* [2004].

## 3. $NO_2$ opacity and the energy balance

$NO_2$ (a brown gas) will block considerable blue and near-UV light. *Reid and McAfee* [1978] pointed out that major ionizing events might easily generate enough opacity to cool the climate, possibly inducing glaciation. Our computational model follows the evolution of various species, and we use it to track the column density of $NO_2$ as a function of latitude, altitude, and time, and its consequent optical depth, as described below.

There are two mitigating effects of the results shown below. The first is that increased transmission of UVB due to $O_3$ depletion will provide some increased heating. This effect is quite small and can be safely ignored for this estimate. The second is that the energy absorbed by the $NO_2$ (which is only generated in significant amounts in the stratosphere) will be re-radiated in the infrared, partially reaching the ground. While a full radiative transfer treatment is beyond the scope of this study, this possibility must be borne in mind considering the results. At most, it could reduce the cooling by a factor of two, as discussed by *Reid et al.* [1978].

The first step in evaluating the degree of this transparency reduction was finding the surface irradiance at any given time in the presence of $NO_2$. This was done by convolving the solar spectral irradiance with the transparency of the Standard US Atmosphere [*ASTM Subcommittee G3.09* 1999] plus $NO_2$. The optical depth was computed using the absorption cross section of $NO_2$ [*Vandaele et al.* 1998] and the column density through the atmosphere at the angle between the Sun direction and a normal to the Earth's surface. This entire integral was then multiplied by the cosine of the angle between the Sun and the surface normal to account for the distribution of energy across the Earth's surface. The irradiance was averaged over the course of a day. This calculation was repeated with the $NO_2$ column density set to zero to obtain the average daily irradiance in the absence of $NO_2$. The relative transparency was taken as the ratio of these two mean irradiance values, and is shown in Figure 1. We show a full year pre-burst, so that the change due to the burst is more obvious.

We have presented the transparency evolution after two different bursts, which bracket the range of effects variation based on latitude and time of year. One burst is at the December solstice at 45° N, the other at the March equinox over the equator. Effects scale with the total fluence of the burst at Earth, although not linearly. Results should be generally insensitive to details of GRB time variation, duration, or detailed spectrum of the GRB within the range of likely variation. Detailed results are shown in the figures, but both models contain global average fluence reductions of order 1% for timescales of years. The December burst shows large reduction, of order 30%, in the polar regions for two years. This is a general feature of non-equatorial bursts, with intensity varying by the burst time of year [*Thomas et al.* 2005b].

Our results are suggestive that climate change may be possible from the several years' reduced sunlight. For example, an estimated reduction of only 0.36% in solar flux during the Maunder Minimum may have caused the "Little Ice Age" in Europe [*Hoyt and Schatten* 1993]. In our simulations, the greatest $NO_2$ buildup takes place in high latitudes, partially persisting during polar summer, which ought to have a greater effect, reducing the melting of ice there. A glaciation accompanied the late Ordovician extinction. Its initiation is not well understood. A drawdown of $CO_2$ in the late Ordovician may have set the stage for a glaciation, [*Patzkowsky et al.* 1997; *Pancost et al* 2003], but may have required a perturbation such as that found here in order to trigger the instability [*Herrmann et al.* 2003, 2004]. Further progress in understanding this process will require detailed climate modeling.

## 4. Rainout of nitrate onto the biosphere

$NO_2$ reacts with hydroxyl to make nitric acid $HNO_3$. This is rained out, which is one of the primary ways the atmosphere returns to equilibrium after our model burst. The global mean surface deposition of nitrate is interesting, of order the amount deposited by lightning over several years, or the amount used in a typical agricultural application. Nitrogen is essential for life, but atmospheric $N_2$ is nearly unavailable to most organisms due to the strength of the nitrogen triple bond. Most biota respond strongly with increased growth rates due to nitrate deposition [*Schlesinger* 1997]. Acids would stress portions of the biosphere, but after immediate titration would act as fertilizer.

In order to compute nitrate deposition, we used $HNO_3$ rainout data as a reasonable measure. Rainout is directly computed by the GSFC 2D atmospheric computational model. Using both $HNO_3$ concentration and rainout coefficients from the atmospheric model, we computed $HNO_3$ rainout rates as a function of time and latitude over a period of twenty years, output every month, giving the cumulative nitrate deposition. The data that we have used as the input for this run are the short-run output data from immediately after the two bursts, as well as a base-line run without a burst. Significant rainout takes at least a few months.

Our rainout rate estimates, and in particular their geographic pattern, must be regarded as only exemplary, since they are based on current empirical data. Rainout is more strongly affected than are ozone depletion and radiation transparency by the configuration of surface features, such as land versus sea, mountains, etc., and overall temperature, which would be different at other times. These results should be viewed as a rough guide to what may be expected. They would also vary according to GRB burst energy, latitude, and season, as do the other computed outcomes.

Figure 2 shows global nitrate rainout rates (flux of N as nitrate) pre and post-burst. Much of the latitude asymmetry in this case is due to the fact that there is presently more rain in the northern hemisphere. This would not be a feature expected at other times. However, we believe our global averages are good guidelines. Our baseline rainout rate is empirical and will also include a small additional component due to biological sources. N fixation by land plants would not be a significant contribution during the Ordovician. However, most biogenic nitrogen is directly deposited on land or in water and not from rainout. The background rate due to lightning constitutes at most 70% of the pre-burst background as shown in this Figure [Schlesinger 1997]. Thus, a burst at the Ordovician or earlier would cause a somewhat greater fractional change in nitrate rainout rate.

The global average annual nitrate rainout is in the December burst (for about five years) almost equal to the total generated by lightning and other non-biogenic sources [*Schlesinger* 1997]. For the other burst, it is about half this amount. Of course, as shown in Figure 2, the deposition is very patchy, basically determined by precipitation. This deposition amount is not large by modern standards, but would constitute an approximate doubling of the nitrate precipitation onto land and shallow water ecosystems at early times. Very little data exists on the nonlinear response of nitrate-starved ecosystems to supplementation *[Schlesinger* 1997, Fig. 12.7], but we could surmise at least a doubling of productivity in such systems for a few years. The additional nitrate would persist in areas where it did not wash away.

Increased productivity is indicated by isotopic excursions in $\delta^{13}C$ at the late Ordovician extinction, coincident with $\delta^{18}O$ indicating lower temperatures. [*Sheehan* 2001; *Brenchley et al.* 2003] In addition, increased productivity could lower atmospheric $CO_2$, providing further positive feedback toward glaciation. We can estimate the size of this effect, using data *from Schlesinger et al.* [1997]. Wetlands typically sequester of order 1000 g C $m^{-2}$ $y^{-1}$ and a bit less that 1% of the Earth surface is contained in such systems. Ignoring land (largely unoccupied) and open ocean (large dilution factor) for the Ordovician, a doubling of wetland productivity would remove about 0.3% of the present $3 \times 10^{18}$ g $CO_2$ per year, we conclude this effect could only be significant with a more powerful GRB or greater nitrate sensitivity during the Ordovician.

## 5 A smoking phase pistol?

The phrase "smoking gun" has been used to describe, for example, the iridium layer associated with the impact that is credited with the end-Cretaceous extinction. Nothing so definite is possible with a GRB, as it is an extremely "clean" event. The only unambiguous prediction is the nitrates, but at less than 1 g $m^{-2}$ net deposition, solubility is likely to have washed away that signal. If, as some argue [*Dermer and Holmes* 2005], a large flux of high energy cosmic rays accompanied the photons, one might see some remnant of atmospheric spallation such as a small $^{36}Ar$ excess. A GRB is definitely too far away for physical transport of such isotopes as $^{244}Pu$ or $^{60}Fe$ associated with nearby supernovae [*Ellis et al* 1996]. Finding them or an iridium layer would falsify the GRB hypothesis.

## 6 Conclusions

In addition to significant ozone depletion lasting several years, a gamma-ray burst within our galaxy at a reasonable distance based on current rate information would generate significant opacity due to $NO_2$. The intensity of such effects depend upon the distance of the burst, as well as the latitude and time of year at which it irradiates the Earth. They are, however, of an interesting magnitude for cooling climate. Whether they are a reasonable perturbation for initiating the late Ordovician glaciation will require detailed climate modeling.

The GRB scenario is also associated with a global deposition of nitrate in the form of dilute $HNO_3$ rain lasting up to a decade. The total amount is comparable to that generated from lightning, the primary non-biogenic source. This could have had a stimulant effect on the biosphere, especially prior to the onset of significant nitrogen-fixing land flora, but the amount may be too small for major effects.

The late Ordovician extinction appears consistent with GRB causation. Such an irradiation seems unlikely to leave detectable residue, but may be falsified by the detection of isotopes associated with supernovae or bolide impacts.

## 6. Acknowledgments

This material is based upon work supported by the National Aeronautics and Space Administration under Grant NNG04GM14G issued through the Astrobiology: Exobiology and Evolutionary Biology Program and supercomputer support from NCSA. LE also acknowledges support from an undergraduate research award of the Honors Program at the University of Kansas.

# 7. References

ASTM Subcommittee G3.09.(1999) *G173-03 Standard Tables for Reference Solar Spectral Irradiances: Direct Normal and Hemispherical on 37° Tilted Surface*, 20pp, ASTM International. Data online at http://rredc.nrel.gov/solar/spectra/am1.5/

Band, D. et al.(1993) BATSE observations of gamma-ray burst spectra. I - Spectral diversity, *Astrophys. J.* 413, 281-292.

Brenchley, P.J., G.A. Carden, L. Hints, D. Kaljo, J.D. Marshall, T. Martma, T. Meidla, and J. Nõlvak (2003) High resolution stable isotope stratigraphy of Upper Ordovician sequences: Constraints on the timing of bioevents and environmental changes associated with mass extinction and glaciation, *GSA Bulletin* 115, 89-104.

Cockell, C.S. (1999) Crises and extinctions in the fossil record—a role for ultraviolet radiation? *Paleobiology* 25, 212-225.

Considine, D.B., A.R. Douglass, and C.H. Jackman(1994) Effects of a polar stratospheric cloud parameterization on ozone depletion due to stratospheric aircraft in a two-dimensional model, *J. Geophys. Res.* 99, 18879-18894.

Dermer, C.D., and J.M. Holmes (2005) Cosmic rays from gamma-ray bursts in the galaxy, *Astrophys. J. Lett*., in press (astro-ph/0504158)
.
Douglas, A.R., C.H. Jackman, and R.S. Stolarski (1989) Comparison of Model Results Transporting the Odd Nitrogen Family With Results Transporting Separate Odd Nitrogen Species*, J. Geophys. Res.* 94, 9862-9872.

Ellis, J., Fields, B.D., and Schramm, D.N(1996) Geological Isotope Anomalies as Signatures of Nearby Supernovae *Astrophys. J.* 470, 1227-1236

Fleming, E.L., C.H. Jackman, R.S. Stolarski, and D.B. Considine (1999) Simulation of stratospheric tracers using an improved empirically based two-dimensional model transport formulation, *J. Geophys. Res.* 104, 23911-23934.

Gensel, P.G. and D. Edwards (2001) *Plants invade the land: Evolutionary and Environmental Perspectives*, Columbia University Press, New York.

Häder, D., Kuma, H.D., Smith, R.C., and Worrest, R. (2003) Aquatic ecosystems: effects of solar ultraviolet radiation and interaction with other climate change factors, *Photochem. Photobiol. Sci.* 2, 39-50.

Herrmann, A.D., M.E. Patzkowsky, and D. Pollard (2003) Obliquity forcing with 8–12 times preindustrial levels of atmospheric $p$CO$_2$ during the Late Ordovician glaciation: *Geology*, 31, 485-488.

Herrmann, A.D., B.J. Haupt, M.E. Patzkowsky, D. Seidov, and R.L. Slingerland ( 2004) Response of Late Ordovician paleoceanography to changes in sea level, continental drift, and atmospheric $p$CO$_2$: Potential causes for long-term cooling and glaciation: *Palaeogeography, Palaeoclimatology,: Palaeoecology*, 210, 385-401.

Hoyt, D.V., and Schatten, K.H. (1993) A discussion of plausible solar irradiance variations, 1700-1992, *J. Geophys. R.* 98, 18895-18906.

Jackman, C.H., A.R. Douglass, R.B. Rood, and R.D. McPeters (1990) Effect of Solar Proton Events on the Middle Atmosphere During the Past Two Solar Cycles as Computed Using a Two-Dimensional Model, *J. Geophys. Res.* 95, 7417-7428.

Jackman, C.H., E.L. Fleming, S. Chandra, D.B. Considine, and J.E. Rosenfield (1996) Past, present, and future modeled ozone trends with comparisons to observed trends, *J. Geophys. Res.* 101, 28753-28767.

Melott, A.L., B.S. Lieberman, L.D. Martin, M.V. Medvedev, B.C. Thomas, J.K. Cannizzo, N. Gehrels, and C.H. Jackman (2004) Did a gamma-ray burst initiate the late Ordovician extinction? *Int. J. Astrobiology*, 3, 55-61.

Mészáros, P. (2001) Gamma-ray bursts: accumulating afterglow implications, progenitor clues, and prospects, *Science* 291, 79-84.

Pancost, R., K.H. Freeman, M.E. Patzkowsky, L, Ainsaar, and T. Martma (2003) Dramatic shifts in biomarker carbon isotopic compositions during the Late Ordovician: Evidence for lower than expected pCO2? EGS - AGU - EUG Joint Assembly, Abstracts from the meeting held in Nice, France, 6 - 11 April 2003, abstract #3735.

Patzkowsky, M.E., L.M. Slupik, M.A. Arthur, R.D. Pancost, and K.H. Freeman (1997) Late Middle Ordovician environmental change and extinction: Harbinger of the Late Ordovician or continuation of Cambrian patterns? *Geology* 25, 911-914.

Reid, G.C., and J.R. McAfee (1978) Effects of intense stratospheric ionisation events, *Nature* 275, 489-492.

Saltzman, M.R., and Young, S.A. (2005) Long-lived glaciation in the Late Ordovician? Isotopic and sequence-stratigraphic evidence from western Laurentia, *Geology* 33, 109-112.

Scalo, J., and J.C. Wheeler (2002) Astrophysical and Astrobiological Implications of Gamma-Ray Burst Properties, Astrophys. J. 566, 723-737.

Schlesinger, W.H.(1997) *Biogeochemistry*, 2$^{nd}$ ed., 588 pp, Academic Press, San Diego.

Sheehan, P. (2001) The late Ordovician mass extinction, *Ann. Rev. Earth. Planet. Sci*. 29, 331-364.

Smith, D.S., Scalo, J., and Wheeler, J.C. (2004) Transport of Ionizing Radiation in Terrestrial-like Exoplanet Atmospheres. *Icarus*, 171, 229-253.

Thomas, B.C., C.H. Jackman, A.L. Melott, C.M. Laird, R.S. Stolarski, N. Gehrels, J.K. Cannizzo, and D.P. Hogan (2005a) Terrestrial ozone depletion due to a Milky-Way gamma-ray burst, *Astrophys. J. Lett.*, 622, L153 (astro-ph/0411284).

Thomas, B.C., C.H. Jackman, A.L. Melott, C.M. Laird, R.S. Stolarski, N. Gehrels, J.K. Cannizzo, and D.P. Hogan (2005b) Gamma-ray bursts and terrestrial ozone depletion: exploration of latitude and seasonal variations, *Astrophys. J.*, submitted.

Thorsett, S.E. (1995) Terrestrial implications of cosmological gamma-ray burst models, *Astrophys. J. Lett.* 444, L53-L55.

Vandaele, A.C., C. Hermans, P.C. Simon, M. Carleer, R. Colin, S. Fally, M.F. Mérienne, A. Jenouvrier, and B. Coquart.(1998) *Measurements* of the NO2 absorption cross-section from 42 000 $cm^{-1}$ to 10 000 $cm^{-1}$ (238—1000 nm) at 220 K and 294 K., *J. Quant. Spectrosc. Radiat. Transfer*, 59, No. 3-5, pp. 171-184. Data online at http://www.oma.be/BIRA-IASB/Scientific/Topics/Lower/LaboBase/Laboratory.html

Waxman, E. (2004) High energy cosmic rays from gamma-ray burst sources: a stronger case, *Astrophys. J.,* 606, 988-993.

Wellman, C.H., Osterloff, P.L., and Mohluddin, U. (2003) Fragments of the earliest land plants *Nature* 425, 282-284.

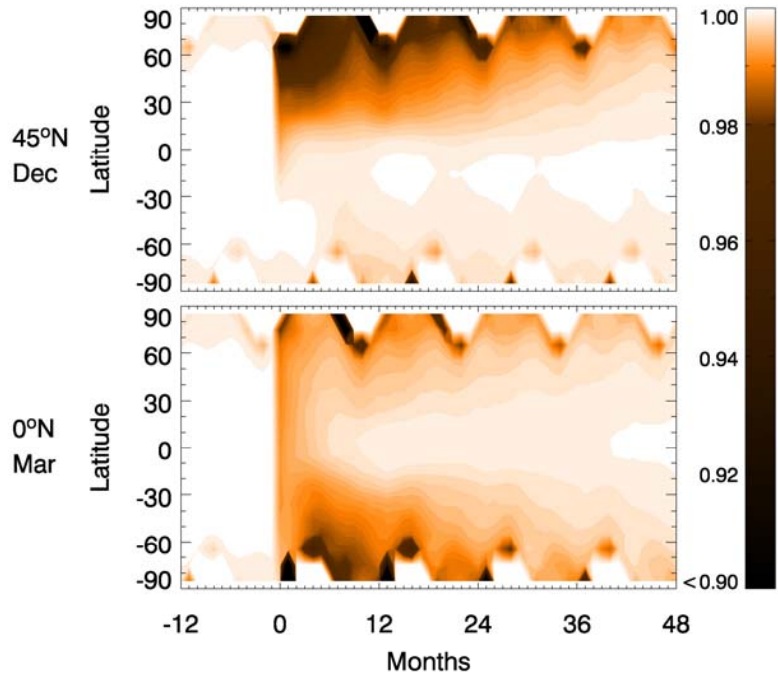

Figure 1: The time development of relative solar fluence reaching the surface for two different GRBs at month 0 at the given time and latitude. Relative fluence is the computed amount divided by that in the absence of $NO_2$ absorption. The value is set to 1 for total polar darkness.

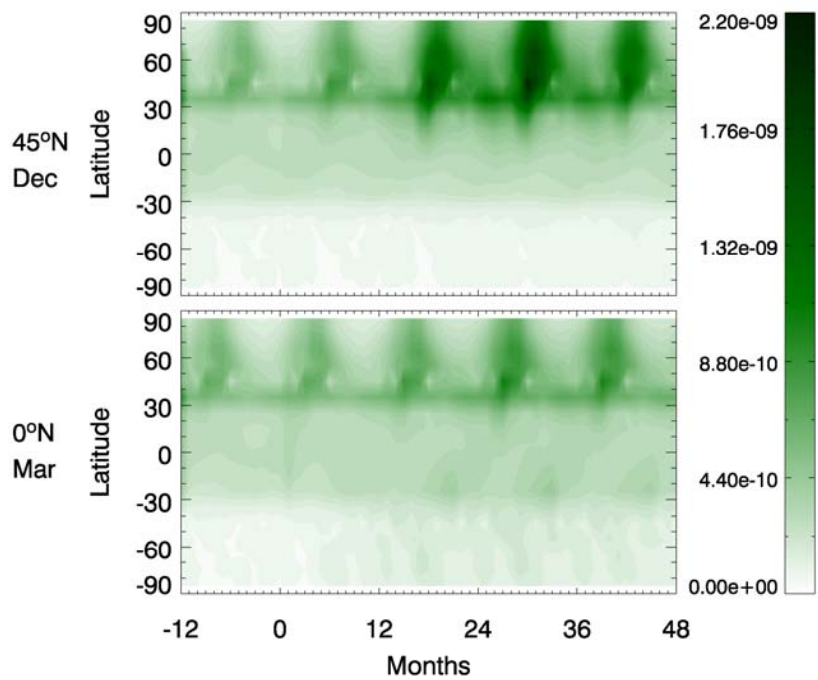

Figure 2: The time development of rainout of N as nitrate in g m$^{-2}$ s$^{-1}$ for the same two bursts. The pre-burst background includes all non-anthropogenic rainout at modern times, and would be somewhat lower prior to terrestrial nitrogen-fixing plants.